\documentclass[runningheads]{svmult}
\usepackage{makeidx}   
\usepackage{graphicx}  
\usepackage{subeqnar}  
\usepackage{multicol}  
\usepackage{cropmark} 
\usepackage{physprbb}  
\begin{document}
\title*{Axino as Cold Dark Matter Candidate}
%
%
%
%
\titlerunning{Axino as CDM}
%
\author{Jihn E. Kim
}

\institute{Division of Physics, Seoul National University, Seoul
151-747, Korea}

\maketitle              

\begin{abstract}
The possibility of the CDM axino is presented. From the early days
of supersymmetry, axino has been considered as hot dark matter and
warm dark matter. But the CDM axino has been considered quite
recently. It is because the low reheating temperature of the
universe is gaining momentum recently from various considerations
in particle cosmology. I present a general introduction on the
CDM's, in particular the axino, the calculation on the axino
density in the universe, and the constraints imposed on the axino
parameters.
\end{abstract}
\def\Oc{\Omega_{\rm CDM}}
\def\OL{\Omega_{\Lambda}}

\section{Introduction}

In this talk, I review the work collaborated with Covi {\it et
al.} \cite{ckr,ckkr} on the possibility of the CDM axino.

Modern cosmology needs dark matter and dark energy in the
universe: $\Omega_{\rm CDM}\simeq 0.3,\ \OL\simeq 0.65$. There are
several particle candidates for the CDM: heavy neutrino in the GeV
range, axion in the $10^{-5}$~eV range, the lightest
superpartner(LSP) in the 100~GeV range, wimpzilla in the
$10^{12}$~GeV range, axino, and other hypothetical heavy
particles.

The particle mass and the interaction strength are provided by
particle physics. In the standard Big Bang cosmology, this
information leads to the decoupling temperature($T_D$) below which
the number of particles in the comoving volume is preserved. But
the needed inflationary period reheats the universe, after a
period of supercoolong, to the reheating temperature $T_R$. If
$T_D>T_R$, the particles are mostly diluted out and the first
guess on its number density is that it cannot serve as dark
matter(DM). But even after the reheating phase, the thermal
production can be effective enough to create sufficient number of
heavy particles to close the universe. It has been most
extensively studied for the case of O(100~GeV)
gravitino\cite{gravitino}. Because of the gravitino problem, we
consider the reheating temperature below $10^9$~GeV.\footnote{ For
the wimpzilla case, the exit phase of the inflation produce the
required density.}

For $T_D<T_R$, the DM candidates are

\begin{itemize}
\item Neutrinos: $G_F$ is the interaction strength.
\item LSP: $\frac{1}{M^2_{\rm SUSY}}$ is the interaction strength.
\end{itemize}

For $T_D>T_R$, the DM candidates are

\begin{itemize}
\item Axion: $\frac{1}{F_a}$ is the interaction srength.
Collective motion of axions below the QCD phase transion
contributes.
\item Axino: $\frac{1}{F_a}$ is the interaction strength. Thermal
and nonthermal productions contribute.
\end{itemize}

\section{Axion and gravitino constraints}

Before discussing the axino CDM possibility, let us briefly review
the cold axions, collectively moving in the universe. The basic
difference of the axion candidate from the other ones is that
axion is boson and it is possible to have a collective motion. The
classical axion potential is extremely flat due to the very tiny
invisible axion mass\cite{inaxion}. Because of the extremely weak
interaction strength, the axion vacuum stays at a point $\langle
a\rangle$ for a long time. The vacuum starts to oscillate when the
Hubble time ($\sim 1/H$) is larger than the oscillation period
($1/m_a$),$ H<m_a $ which occurs when the temperature is about
1~GeV\cite{review}. We understand that the axion is the
pseudo-scalar field, having the interaction only through the
anomaly
\begin{equation}
\frac{1}{32\pi^2}\frac{a}{F_a}F_{\mu\nu}\tilde F^{\mu\nu}\equiv
\frac{a}{F_a}\left\{F\tilde F\right\}
\end{equation}
where $\tilde F_{\mu\nu}$ is the dual field strength $(1/2)
\epsilon_{\mu\nu\rho\sigma}F^{\rho\sigma}$. We stress that there
should not exist any other term in the axion potential. This kind
of nonrenormalizable interaction can arise in several different
ways:
\begin{itemize}
\item From fundamental interaction such as in
superstring\cite{witten}: $F_a\sim$ Planck mass.
\item From composite models\cite{composite}: $F_a\sim$ compositeness
scale.
\item From renormalizable theories\cite{peccei}: A global symmetry
can have a gluon anomaly. If this global symmetry is spontaneously
broken, there arises a Goldstone boson coupling to the gluon
anomaly. $F_a\sim$ the global symmetry breaking scale.
\end{itemize}
Then the current axion energy density is estimated to be\cite{pww}
\begin{equation}
\rho_a(T_\gamma)=m_a(T_\gamma)n_a(T_\gamma)\simeq
2.5\frac{F_a}{M_P}\frac{F_am_a}{T_1}T_\gamma^3\left(
\frac{A(T_1)}{F_a}\right)^2.
\end{equation}
If $F_a$ is large($>10^{12}$~GeV), the axion energy dominates the
current energy density in the universe. Since the energy density
is proportional to the number density, it behaves like a CDM. On
the other hand, if $F_a$ is small then the axion interaction is
relatively strong and too much axions are produced in the core of
stars. SN1987A restricts its lower bound around $10^9$~GeV,
leading to the axion window
\begin{equation}
10^9~{\rm GeV}<F_a<10^{12}\ {\rm GeV}.
\end{equation}

The gravitino produced thermally after inflation decays very late
in the cosmic time scale($>10^3$~s), and can dissociate the light
nuclei by its decay products\cite{weinberg}. Inflation was used to
dilute the primordial gravitinos, but thermal production of
gravitinos are troublesome in cosmology if it
exceeds\cite{gravitino}
\begin{equation}
T_R>10^9\ {\rm GeV}.
\end{equation}
Therefore, in supersymmetric theories we must consider relatively
small reheating temperature after inflation.

\section{Axino}

Therefore, in supersymmetric extension of the axion model, axion
can be a CDM candidate but it is produced very late. Its
supersymmetric partner is axino, and the reheating temperature
$T_R$ must be smaller than $T_D$. The axion $a$ and axino $\tilde
a$ are gathered in a supermultiplet,
\begin{equation}
\Phi=\frac{1}{\sqrt{2}}(s+ia)+\sqrt{2}\tilde a\theta+F_\Phi
\theta\theta
\end{equation}
where $s$ is the scalar partner of axion, the saxion. The axion
couples to the gluon anomaly through $(1/F_a) \{F\tilde F\}$. It
is known that the potential arising from this gluon interaction
settles $\langle a\rangle$ at zero, which is the Peccei-Quinn
mechanism.

For axino to be CDM, it must be stable or practically stable.
Without the $R$-parity conservation, this cannot happen. Thus, we
require a practical {\it R-parity conservation}. For axino to be
the LSP, it must be lighter than the lightest neutralino whose
mass is expected to be around 100~GeV. Thus, an estimate of the
axino mass is of prime importance for a CDM axino. Note that there
is no theoretical upper bound on the axino mass.

Since axion is almost massless, one might expect that the axino
and saxion are almost massless in the first approximation.
However, the saxion obtains a soft mass term below SUSY breaking
scale. It is like the SM SUSY scalars. So the axino mass is
intimately reated to the SUSY breaking scenario also. Actually, it
is known that the axino obtains a substantial
mass\cite{polchinski}. For a specific model, including the soft
SUSY breaking terms, consider
\begin{equation}
V=|f|^2(|S_1|^2+|S_2|^2)|Z|^2+(A_1fS_1S_2Z-A_2fF_a^2Z+{\rm h.c.})
\end{equation}
where $Z,S_1,$ and $S_2$ are the SM singlets, and the
supersymmetric term comes from the superpotential
$W=fZ(S_1S_2-F_a^2)$. Since $\langle S_i\rangle$ is of order
$F_a$, $Z$ is of order the $A$ term. Thus, the fermionic partners
have the mass matrix of the form,
\begin{equation}
\matrix{S_1\cr S_2\cr Z\cr}\ \ \left(\matrix{0\ \ \ \ m_{\tilde
a}\ fF_a \cr m_{\tilde a}\ \ \ \ 0\ \ fF_a\cr fF_a\ fF_a\ \ 0
\cr}\right)
\end{equation}
where $m_{\tilde a}=f\langle Z\rangle$. One eigenvalue is the
axino mass $\sim m_{\tilde a}$, and the other masses are of order
$F_a$. As seen from this example, the axino mass is basically a
free parameter, but is expected to be somewhat smaller than the
naive SUSY breaking scale due to the coupling $f$\cite{yamaguchi}.
But in some models, the axino mass can be much smaller than the
SUSY breaking scale. Take a superpotential
$W'=fZ(S_1S_2-X^2)+(\lambda/3)(X-M)^3$ with one more singlet $X$
which carries a vanishing PQ charge. This superpotential is much
more complicated to analyze, but still we can show that $m_{\tilde
a}= O(A-2B+C)+O(m^2_{3/2}/F_a)$. For the standard pattern of soft
terms, $B=A-m_{3/2}, C=A-2m_{3/2}$. Then, the axino mass is of
order keV. Thus, even the tree level axino mass needs the
knowledge on the full superpotential\cite{nilles}. Therefore, we
can consider the axino mass a free parameter, and we restrict our
discussion in the region where axino is the LSP, which is the most
probable choice with the PQ symmetry.

The universal axion coupling is to the anomaly term. Therefore,
the axino coupling to the gauge multiplet is the most important
couplings. For the fermions, only the coupling to the top quark is
important. We consider all these couplings when they become
appropriate.

\section{Axino density in the universe}

The axino decoupling temperature is
\begin{equation}
T_D\simeq 10^{10}\ {\rm GeV}\
\left(\frac{F_{a,11}}{N_{DW}}\right)\left(\frac{0.1}{\alpha_s}
\right)^3
\end{equation}
where $N_{DW}$ is the domain wall number in axion models and
$F_{a,11}$ is the axion decay constant in units of $10^{11}$~GeV.
For $T_D<T_R$, at $T_D$ the number density is determined. For
axinos not to close the universe, it should not exceed
12.8~eV~$g_*(T_D)/g_{eff}$ where $g_*=915/4, g_{eff}=1.5$, which
was used for the O($<2$~keV) warm DM axino by Rajagopal {\it et
al.}\cite{warmdm}. But in our $T_R<T_D$ case, due to the gravitino
problem\cite{gravitino}, we do not consider this region seriously.
On the other hand, we can consider much heavier axinos. The
estimation of the axino density in the universe is considered for
the thermal case and for the non-thermal case(by the neutralino
decay).

For the thermal production, we solve the Boltzmann equation,
\begin{equation}
\frac{dn_{\tilde a}}{dt}=\sum_{i,j}\langle \sigma(i+j\rightarrow
\tilde a+\cdots)v_{rel}\rangle n_in_j+\sum_{i}\langle
\Gamma(i\rightarrow\tilde a+\cdots)\rangle n_i.
\end{equation}
The axino yield, the number density divided by the entropy
$s=(2\pi^2/45)g_{s*}T^3$, is split into two pieces one from the
scattering process and the other by the decay, $Y^{TP}_{\tilde
a}=\sum_{i,j} Y^{scatt}_{i,j}+\sum_{i}Y_i^{dec}$ where
$$
Y_{i,j}^{scatt}=\int_0^{T_R}dT\frac{\langle\sigma(i+j\rightarrow
\tilde a+\cdots)\rangle n_in_j}{sHT},\ \
Y_i^{dec}=\int_0^{T_R}dT\frac{\langle\Gamma(i\rightarrow\tilde
a+\cdots)\rangle n_i}{sHT}.
$$
For the scattering, we consider ten important processes shown in
Table 1.
\begin{table}
\caption{The cross sections for the different axino thermal
production channels involving strong interactions. Masses are
neglected except for the plasmon mass $m_{eff}$.}
\begin{center}
\renewcommand{\arraystretch}{1.4}
\setlength\tabcolsep{5pt}
\begin{tabular}{lclccc}
\hline\noalign{\smallskip} n & Process & $\overline{\sigma}_N$ &
$n_{\rm spin}$ & $n_F$ & $\eta_1\eta_2$\\
\noalign{\smallskip} \hline \noalign{\smallskip}
A & $g^a+g^b\rightarrow\tilde a +\tilde g^c$ & $\frac{1}{8}
|f_{abc}|^2$ & 4 & 1 & 1 \\
B & $g^a+\tilde g^b\rightarrow\tilde a +g^c$ & $\frac{5}{16}|f_{abc}|^2
[\ln(s/m^2_{eff})-\frac{15}{8}]$ & 4 & 1 & $\frac{3}{4}$ \\
C & $g^a+\tilde q_k\rightarrow\tilde a +q_j$ & $\frac{1}{8}|
T^a_{jk}|^2$ & 2 & $N_F\times 2$ & 1 \\
D & $g^a+q_k\rightarrow\tilde a +\tilde q_j$ & $\frac{1}{32}
|T^a_{jk}|^2$ & 4 & $N_F\times 2$ & $\frac{3}{4}$ \\
E & $\tilde q_j+q_k\rightarrow\tilde a +g^a$ &
$\frac{1}{16}|T^a_{jk}|^2$
& 2 & $N_F\times 2$ & $\frac{3}{4}$ \\
F & $\tilde g^a+\tilde g^b\rightarrow\tilde a +\tilde g^c$ &
$\frac{1}{2}|f_{abc}|^2[\ln(s/m^2_{eff})
-\frac{29}{12}]$ & 4 & 1 & $\frac{3}{4}\ \ \frac{3}{4}$\\
G & $\tilde g^a+q_k\rightarrow\tilde a +q_j$ &
$\frac{1}{4}|T^a_{jk}|^2[\ln(s/m^2_{eff})
-2]$ & 4 & $N_F$ & $\frac{3}{4}\ \ \frac{3}{4}$ \\
H & $\tilde g^a+\tilde q_k\rightarrow\tilde a +\tilde q_j$ &
$\frac{1}{4}|T^a_{jk}|^2[\ln(s/m^2_{eff})
-\frac{15}{8}]$ & 2 & $N_F\times 2$ & $\frac{3}{4}$ \\
I & $q_j+\bar q_k\rightarrow\tilde a +\tilde g^a$ & $\frac{1}{24}
|T^a_{jk}|^2$ & 4 & $N_F$ & $\frac{3}{4}\ \ \frac{3}{4}$ \\
J & $\tilde q_j+\tilde q_k\rightarrow\tilde a +\tilde g^a$
& $\frac{1}{24}|T^a_{jk}|^2$ & 1 & $N_F\times 2$ & 1 \\
\hline
\end{tabular}
\end{center}
\label{Tab1.1a}
\end{table}
On the other hand, the non-thermal production is basically by the
decay process of the very long lived neutralino\cite{ckr}. These
contributions are shown in Fig. 1. For a low reheating
temperature, the thermal production is not effective. In Fig. 1,
the bulge at the lower left corner is the non-thermal production.
\begin{figure}[b]
\begin{center}
\includegraphics[width=.95\textwidth]{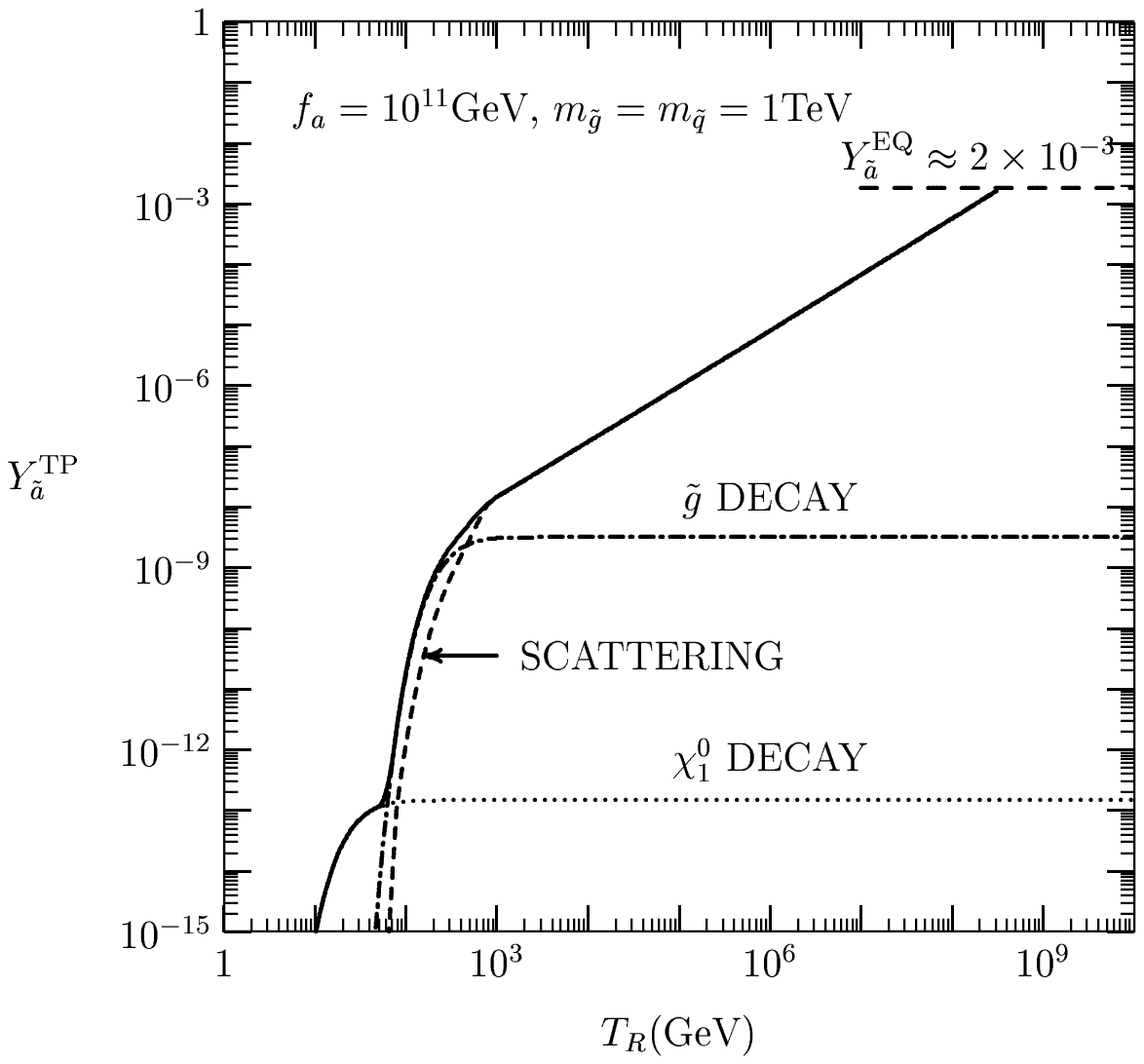}
\end{center}
\caption[]{$Y_{\tilde a}$ as a function of $T_R$ for
$F_a=10^{11}$~GeV and $m_{\tilde q}=1$~TeV.} \label{eps1.1}
\end{figure}
In Fig. 1, we show for $F_a=10^{11}$~GeV so that the cold axions
are not the dominant component of CDM.

For the calculation of the non-thermal production, we consider
LOSP and NLSP where LOSP is the lightest superpartner among
ordinary SM particles and NLSP is the next-to-LSP or the second
lightest superpartner. It is most likely that the LOSP is the
lightest bino-like neutralino and NLSP can be LOSP or gravitino or
something else. For a bino-like LOSP $\chi$, the decoupling
temperature of the LOSP is $T_D\simeq m_\chi/20$. At $T<T_D$ and
$\Gamma_\chi\ll H$, the LOSP decays during the radiation dominated
era to produce axinos through $\chi\rightarrow \tilde a+\gamma$ to
yield\cite{ckkr}
\begin{equation}
Y_\chi(T)\simeq Y_\chi^{EQ}(T_D)\exp \left(
-\int_T^{T_D}\frac{dT'}{T^{\prime 3}}\frac{m_\chi^3\langle
\Gamma_\chi\rangle_{T'}}{H(m_\chi)} \right).
\end{equation}
Therefore, we have a simple expression,
\begin{equation}
\Omega_{\tilde a}=\frac{m_{\tilde a}}{m_\chi}\Omega_\chi.
\label{number}
\end{equation}
The neutralino decays at the cosmic time
\begin{eqnarray}
\tau(\chi\rightarrow\tilde\gamma)=0.33\ {\rm s}\
\frac{1}{C^2_{aYY}Z^2_{11}}\left(\frac{1/128}{\alpha_{em}^2}\right)^2
\left(\frac{F_a/N_{DW}}{10^{11}\ {\rm
GeV}}\right)^2\left(\frac{100\ {\rm
GeV}}{m_\chi}\right)^3\left(1-\frac{m^2_{\tilde
a}}{m^2_\chi}\right)^{-3}\nonumber
\end{eqnarray}
Depending on the models, the decay modes can be different. If
there are more channels for the $\chi$ decay, then it will give an
even more efficient implementation of the non-themal production of
axino through the neutralino decay.


\section{Constraints}

We must impose the conditions that (1) not too much axino energy
density at the time of Big Bang nucleosynthesis(BBN), (2) the BBN
not spoiled by $\chi$ decay producing the SM particles, and (3)
axinos becoming cold before the galaxy formation era.

Note in passing that if $m_{\tilde a}<m_{3/2}< m_\chi$, i.e. for
the gravitino NLSP then the gravitino problem disappears since it
decays to axion and axino\cite{asaka}. On the other hand, if
$\chi$ is the NLSP, the gravitino problem is
present\cite{gravitino}. In this case, the non-thermally produced
axinos are estimated to give Eq.~(\ref{number}).
\begin{figure}[b]
\begin{center}
\includegraphics[width=.95\textwidth]{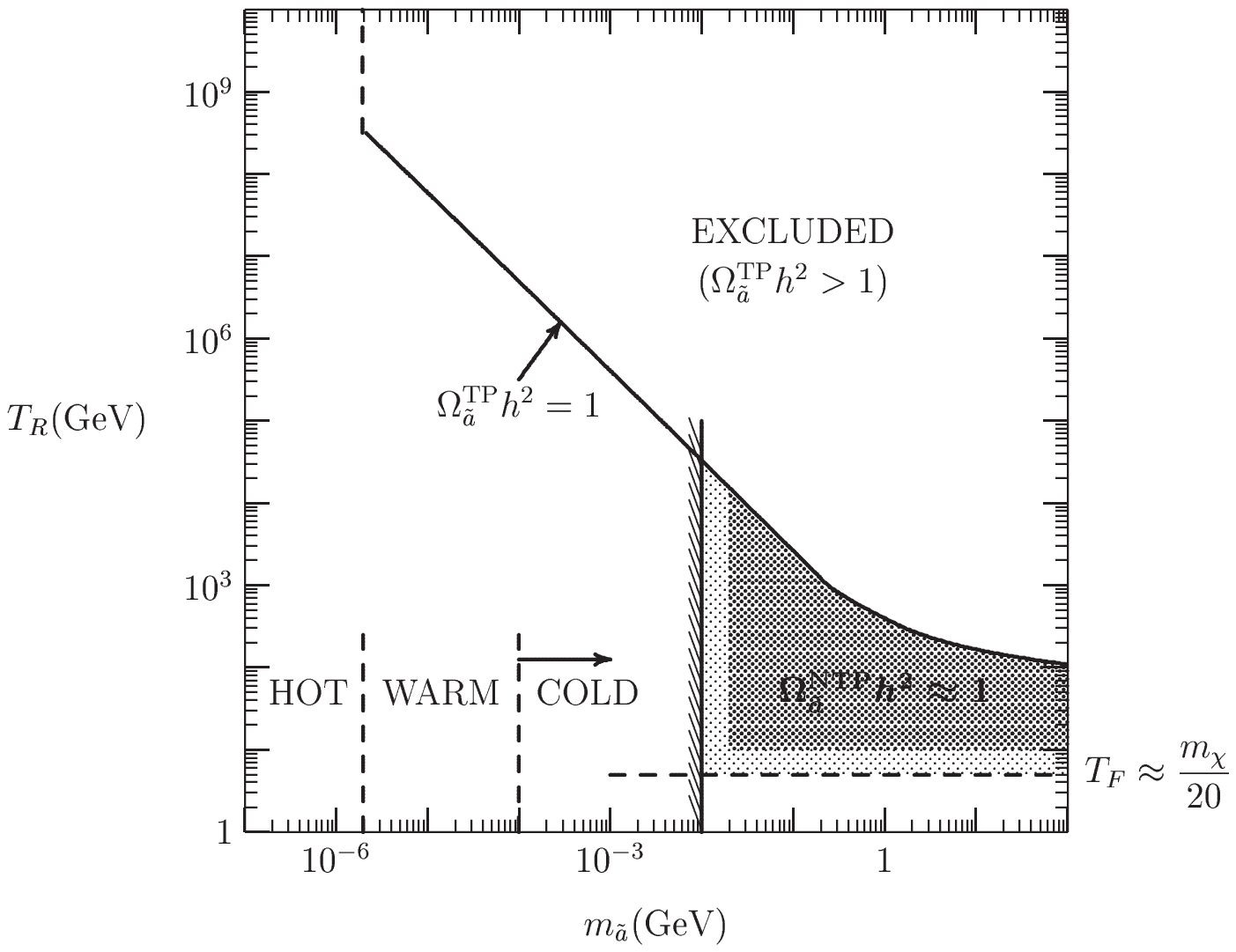}
\end{center}
\caption[]{Constraints on $T_R$ and axino mass.} \label{eps1.1}
\end{figure}
In Fig. 2, we include all these constraints and plot the
restriction on the reheating temperature versus axino mass. This
figure is not changed even if there is no gravitino problem
because of the gravitino NLSP. The shaded region is where the
axino can be a CDM possibility. This happens O(GeV) axino.

Note that for a high reheating temperature the thermal production
contribution dominates. In the figure, there is the solid line
denoting the critical energy density by axinos. Even if the
reheating temperature is below the critical energy density line,
there still exists a axino CDM possibility by non-thermally
produced axinos\cite{ckr}. Note that non-thermally produced axinos
can be CDM for relatively low reheating temperature($<10$~TeV) for
which axino mass should be
\begin{equation}
10\ {\rm MeV}<m_{\tilde a}<m_{\chi}\ \ :\ \ {\rm non-thermal\
axinos\ as\ CDM\ possibility}
\end{equation}
The shaded region corresponds to the MSSM models with the
constraint $\Omega_\chi h^2<10^4$ but still allowable axino energy
density. One can choose 30 \% CDM in this region. However, if we
restrict further that all SUSY mass parameters are below 1~TeV,
then we have $\Omega_\chi h^2<10^2$. For a sufficient axino energy
density for the axino CDM, then we have
\begin{equation}
m_{\tilde a}\ge 1~{\rm GeV}
\end{equation}
but less than the neutralino mass.

\section{Detection possibility of CDM axino}

If $R$-parity is exact, then the CDM axino cannot decay, and there
is no way to prove the CDM axino possibility. If there is any
chance to prove the CDM axino by observation, then the axino must
decay\cite{hbkim}. For it to constitute most of CDM in the
universe, its lifetime must be of order the age of the universe.
Since it is assumed to be the LSP, the $R$-parity must be feebly
broken. Then the CDM axino can decay.

For the simplicity of discussion, let us consider the bilinear
terms only
\begin{equation}
W=\mu_\alpha L_\alpha H_2
\end{equation}
where $\alpha(=1,2,3)$ is the flavor index, $\mu_\alpha=O$(eV) is
the constraint from the neutrino mass bound. This bound applies to
the heaviest neutrino, presumably $\nu_\tau$. In this case $\mu_3$
is bounded as
\begin{equation}
|\mu_3|\le M^{1/2}_{\tilde H_{2,TeV}}\ {\rm MeV}.
\end{equation}
With the $R$-parity violation, we expect the decays
\begin{eqnarray}
\tilde a\rightarrow \nu+\gamma({\rm or}\ l^+l^-),\nonumber\\
\tilde a\rightarrow\nu+a\\
\tilde a\rightarrow \tau^++\pi^-.\nonumber
\end{eqnarray}
In fact, the $\nu\gamma$ mode is most important in cosmology. For
the cosmic importance of the decay, we need to know the lifetime.
Let us try phenomenological Lagrangians,
\begin{equation}
\epsilon_0\phi\tilde a\psi,\ \ \
i\epsilon_1\frac{\alpha_{em}}{F_a}F^{\mu\nu}\tilde a
\gamma_5[\gamma_\mu,\gamma_\nu]\psi,
\end{equation}
for the final pseudo-scalar and photon, respectively. Then, for
each case the lifetime can be estimated. In particular, to the
photon mode
\begin{equation}
\tau_{\tilde a}=\frac{4\times 10^4}{\epsilon_1^2} n_\psi
m^{-3}_{\tilde a,GeV}F^2_{a,12}P_1^{-1}[{\rm s}]
\end{equation}
where $P_1$ is the phase space factor. $\epsilon_1$ is known if
the $R$-parity violation is given. Since the most stringent bound
comes from the diffuse gamma ray, we focus on this mode. The
observed gamma ray flux is
\begin{equation}
\frac{F_\gamma}{d\Omega}\le (10^{-3}-10^{-5})E^{-1}_{GeV}{\rm
cm}^{-2}{\rm sr}^{-1}[{\rm s}^{-1}].
\end{equation}

If $t_{rec}<\tau_{\tilde a}<t_0$ where $t_0$ is the age of the
universe, most axinos have decayed by now and the flux has a peak
at $E_0=(m_{\tilde a}/2)(\tau_{\tilde a}/t_0)^{2/3}$. Since the
photons from axino decay should not exceed the observed flux, we
obtain
\begin{equation}
\tau_{\tilde a}<10^{-10}t_0\Omega^{3/2}_{\tilde a}h^3
\end{equation}
which is in conflict with $\tau_{\tilde a}> t_{rec}\sim
10^{13}$~s.

If $\tau_{\tilde a}>t_0$, a similar consideration gives the axino
lifetime larger than $4\times 10^{24-26}\Omega_{\tilde a}h$~s.
This translates to the $\mu_3$ bound,
\begin{equation}
\mu_3<10^{2-3}\ {\rm eV}.\label{mu3}
\end{equation}
If $\mu_3$ in (\ref{mu3}) is taken, the idea of neutrino mass
generation by $R$-parity violation is not enough. Thus, the
unstable CDM axino is not viable with the neutrino mass generation
by $R$-parity violation. Still, the CDM axino decay can be
detected if the parameters lie in the region given above.

\section{Conclusion}
In conclusion, we discussed
\begin{itemize}
\item
Supersymmetry is the solution of the gauge hierarchy problem, and
the PQ symmetry with a very light axion with $10^9\ {\rm
GeV}<F_a<10^{12}$~GeV is the solution of the strong CP problem.
This leads to the axino which can be the LSP.
\item
The gravitino problem must be considered in this case, with
$T_R<10^9$~GeV.
\item
Due to the much more stronger axino production than the gravitino
production, $T_R$ must be much smaller than the one given by the
gravitino consideration. It can be in a multi TeV region.
\item
Thermal and nonthermal production of axinos are possible. In this
case, O(GeV) axino CDM possibility exists.
\item
Finally, the attempt to explain the neutrino mass via the
$R$-parity violation mechanism is inconsistent with the CDM axino.
However, in this case there is a room for the detection of axino
decay debris.
\end{itemize}

\vskip 0.5cm\noindent {\it Acknowledgments}: This work is
supported in part by the BK21 program of Ministry of Education and
by the KOSEF Sundo grant.

\end{document}